# Thermo-Responsive Toughening in LCST-type Hydrogels: Comparison between Semi-Interpenetrated and Grafted Networks

*Hui Guo*[*,1,2], *Nicolas Sanson*[1,2], *Alba Marcellan*[1,2] *and Dominique Hourdet*[*,1,2]

[1]Laboratoire Sciences et Ingénierie de la Matière Molle, CNRS, PSL Research University, ESPCI Paris, 10 rue Vauquelin, F-75005 Paris, France.

[2]Laboratoire Sciences et Ingénierie de la Matière Molle, Université Pierre et Marie Curie, Sorbonne-Universités, 10 rue Vauquelin, F-75005 Paris, France.

Corresponding authors :

*hui.guo@espci.fr

*dominique.hourdet@espci.fr

# Abstract

Hydrophilic and LCST polymer chains, poly(*N,N*-dimethylacrylamide) (PDMA) and poly(*N*-isopropylacrylamide) (PNIPA), were combined in semi-interpenetrated architectures to investigate their responsive properties in swollen isochoric conditions comparatively to grafted network structures. Using equal weight fractions of PDMA and PNIPA, semi-IPN designed with opposite topologies have demonstrated a thermoresponsive behavior with very different structure/properties relationships as investigated by calorimetry, swelling experiments, tensile tests and 2D neutron scattering at rest and under deformation. In the case of the PDMA network interpenetrated by linear PNIPA chains, the phase transition of PNIPA gives rise to the formation of large microdomains, loosely percolating the PDMA network. Above the transition, the enhancement of the mechanical properties remains low in terms of elastic modulus and fracture energy. Conversely, the opposite topology, with PDMA chains interpenetrating the crosslinked PNIPA network, brings a large improvement of the mechanical properties at high temperature with a 10-fold increase of the modulus and very high fracture energy. By comparison with grafted hydrogels of similar composition and different topologies, the impact of the primary structure over the phase-separated morphology and the resulting mechanical properties was clearly highlighted.

# Introduction

Polymer hydrogels, that are three-dimensional networks swollen in water, have received considerable attention during the last decades due to their high potential as soft materials, water containers, texture modifiers or mechanical transducers with applications in various fields like food, agriculture, water treatment, cosmetics, oil recovery and microfluidics. They are actually used in many bioapplications[1-4] such as superabsorbents, contact lenses, dressings, drug delivery systems or scaffolds for tissue engineering, with a very special focus today on biomedicine due to their similarities with biological tissues.[5] In this context many efforts have been made during the last 15 years to reinforce the mechanical properties of these soft and wet materials which were intrinsically weak. Today one can find two main strategies either with the idea to homogenize the stress distribution within the network with the development of slide-ring gels[6] and ideal tetra-PEG gels[7] or to develop dissipation mechanisms by introducing sacrificial bonds, either covalent or physical,[8-10] as in ionic,[11, 12] hydrophobic,[13-16] nanocomposite and hybrid hydrogels.[17-19] Although physical interactions are reversible in nature, they cannot be finely controlled and the synthesis of strongly interacting networks like polyampholytes[12] or hydrogels containing large amount of hydrophobic groups require several step of synthesis and/or solvent exchange.

Recently, we propose to finely tune these secondary interactions within the hydrogel by playing with the thermodynamic properties of polymers in aqueous media and for that purpose we selected poly(*N*-isopropylacrylamide) (PNIPA) for its well-known LCST behavior.[20] As the phase transition of PNIPA that occurs above 32 °C goes hand in hand with a large volume transition,[21] we introduced a hydrophilic counterpart inside the hydrogel in order to maintain a high level of swelling in the segregated regime; i.e. well above the LCST of PNIPA. This has been successfully demonstrated with grafted architectures prepared with equal weight fractions of PNIPA and poly(*N*,*N*-dimethylacrylamide) (PDMA) and high amount of water, almost 85 wt% in the preparation state. By scanning the temperature from 20 to 60 °C, it was shown that grafted hydrogels were able to keep their high original swelling in the whole range of temperature and to develop in isochoric conditions a strong and reversible thermo-toughening behavior with a 10-fold increase of the elastic modulus above the phase transition of PNIPA. From the comparison between hydrogels designed with the same composition but opposite topologies, a crosslinked PNIPA grafted with PDMA side-chains and a crosslinked PDMA grafted with PNIPA side-chains (see GPN-D and GPD-N, respectively, in **Fig.1a**), it was shown that the gel prepared with crosslinked PNIPA was much more resistant to fracture propagation even if initially the two hydrogels display a very similar thermo-toughening behavior at low deformation.[20] From complementary experiments performed above the phase transition of PNIPA by small angle neutron scattering, the difference in mechanical reinforcement was attributed to the bicontinuous structure developed by the GPN-D hydrogel in the segregated regime compared to the micellar morphology of GPD-N.[22]

This responsive strategy allows not only to investigate how the phase separation will reversibly modify the bulk properties of the hydrogel, typically the mechanical behavior, but it is also an interesting way to reach in a controlled manner the strong segregation regime without additional steps of chemistry and/or solvent exchange.

Considering these results, the aim of the present work was to explore new topologies for thermo-toughening hydrogels and for that purpose we developed semi-interpenetrated networks (semi-IPN). With the idea to compare their properties with previous grafted hydrogels, we keep the same preparation conditions with the same weight fractions of PDMA, PNIPA and water and we also designed two semi-IPN architectures with opposite topologies in order to investigate the impact of the phase transition of linear PNIPA embedded into a crosslinked PDMA matrix or the opposite, i.e. a cross-linked PNIPA matrix swollen by PDMA linear chains (see GPD/PN and GPN/PD in **Fig.1b**).

The structure and macroscopic properties of these new semi-IPN hydrogels have been investigated as a function of temperature by swelling experiments, differential scanning calorimetry, rheology, tensile tests and small angle neutron scattering at rest and under deformation. This comprehensive set of analyses will be compared with previous data obtained with grafted hydrogels in order to give a broader overview of the structure/properties relationships.

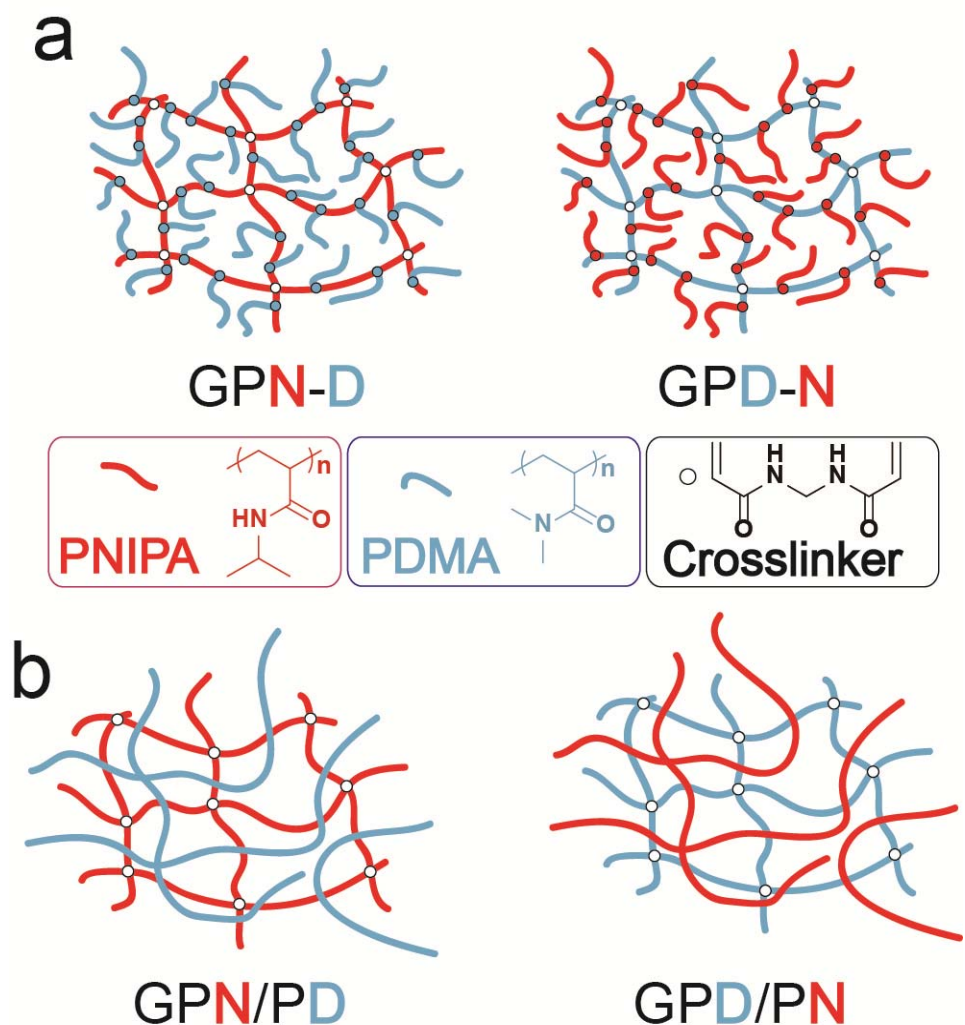

**Fig. 1.** Schematic representation of thermo-responsive hydrogels with opposite PDMA and PNIPA topologies: **(a)** grafted networks (GPN-D and GPD-N) and **(b)** semi-IPN (GPN/PD and GPD/PN) architectures.

# Experimental section

*Materials*

*N*-isopropylacrylamide (NIPA), *N,N*-dimethylacrylamide (DMA), potassium peroxodisulfate (KPS), ammonium persulfate (APS), sodium metabisulfite (SBS), *N,N'*-methylenebisacrylamide (MBA) and tetramethylethylenediamine (TEMED) obtained from Aldrich were used as received. All organic solvents were analytical grade and water was purified with a Millipore system combining inverse osmosis membrane (Milli RO) and ion exchange resins (Milli Q) for synthesis and purification.

*Synthesis of linear polymers*

Linear polymer chains were synthesized by radical polymerization using APS/SBS as initiators. In a three-necked flask, 100 mmol of monomers (NIPA or DMA) was dissolved in 100 mL of water and the solution was deoxygenated for 1 h with nitrogen bubbling. The redox initiator, APS (91.2 mg, 0.4 mmol) and SBS (38.0 mg, 0.2 mmol), were separately dissolved and deoxygenated in 10 mL of water prior addition into the monomer solution. The reaction was allowed to proceed in an ice bath and SEC was used to monitor the polymerization process. After stirring overnight, almost full conversion was achieved, and polymers were purified by dialysis against pure water (membrane cut-off=50 kDa) during one week and recovered by freeze-drying. The linear polymers were obtained with a yield of about 75 wt% for PNIPA and 71 wt% for PDMA. The absolute values of average molar mass and dispersity, characterized by SEC equipped with three detectors in line (refractometry, viscometry and light scattering), are $M_n$=290 kg/mol (Đ=1.5) for PNIPA and 161 kg/mol (Đ=1.4) for PDMA.

The syntheses of PDMA and PNIPA macromonomers were performed as previously described[22, 23] using a telomerization process with cysteamine hydrochloride followed by modification of the amino end-group with acrylic acid in order to get the vinyl function. From previous works performed on similar systems,[24] it was shown that the functionalization of telomers with amino end-group is almost quantitative, at least greater than 90 %, as well as the subsequent vinyl modification of amino-end groups due to the large excess of activated acrylic acid per amino end-groups ([COOH]/[NH$_2$]=15). The absolute characterization of macromonomers by size exclusion chromatography gives the following molar masses: $M_n$=36 kg/mol (Đ=1.4) for PNIPA, $M_n$=39 kg/mol (Đ=1.3) for PDMA and $M_n$=89 kg/mol (Đ=1.5) for PDMA$_{long}$.

*Synthesis of hydrogels*

Monomer, cross-linker (stock solution of MBA) and linear polymer chains (or macromonomers), were initially dissolved in a given amount of water under nitrogen bubbling; the temperature of the reactor being controlled with an ice bath. Stock solutions of KPS and TEMED were separately prepared under nitrogen atmosphere and after 30 min aliquots were added into the reaction medium. After fast mixing (2 min), the final formulation was rapidly transferred between glass plates of 2 mm width under nitrogen atmosphere and the reaction was left to proceed overnight in the fridge (4 °C). The resulting hydrogels were then cut with a die-cutter of rectangular or round shape and directly used for DSC and swelling experiments or stored into incompatible paraffin oil before mechanical testing in order to avoid any change in hydrogel composition induced by swelling or drying. For both topologies, the weight ratio "monomer/linear polymer/water" was kept constant to 1/1/10 (corresponding to an initial swelling ratio $Q_0$=6), and the molar ratio "monomer/MBA/KPS/TEMED" was set equal to 100/0.1/1/1. The hydrogels are denoted **GPN** or **GPD**, according to the polymer used for the crosslinked matrix (**N** for PNIPA and **D** for PDMA), and the nomenclature GPN/PD and GPN-D (and vice versa) will be used for semi-IPN and grafted architectures, respectively. Two semi-IPN and three grafted networks have been used in the present study and the details of their formulation are reported in **Table 1**.

**Table 1** Formulation of thermo-responsive semi-IPN and grafted hydrogels

|  | Monomer g | MBA mg | Polymer g | $M_n$ kg/mol | Đ | KPS mg | TEMED mg | Water g |
|---|---|---|---|---|---|---|---|---|
| GPN/PD | 1 | 1.36 | 1[a] | 161[a] | 1.4 | 23.9 | 10.3 | 10 |
| GPD/PN | 1 | 1.56 | 1[a] | 290[a] | 1.5 | 27.3 | 11.7 | 10 |
| GPN-D | 1 | 1.36 | 1[b] | 39[b] | 1.3 | 23.9 | 10.3 | 10 |
| GPN-D$_{long}$ | 1 | 1.36 | 1[b] | 89[b] | 1.5 | 23.9 | 10.3 | 10 |
| GPD-N | 1 | 1.56 | 1[b] | 36[b] | 1.4 | 27.3 | 11.7 | 10 |

[a]linear polymer and [b]macromonomer

*Gel composition and swelling experiments*

The samples, originally in their preparation state ($Q_0$), were weighed and placed at a given temperature in a large excess of pure water. Water was changed twice every day during four days. The swollen gels were weighed over time ($m_t$) and the swelling ratio ($Q$) were calculated as $Q = m_t/m_d$, $m_d$ being the theoretical dry weight assuming 100% conversion. Finally, after 4 days of equilibrium, the gels were dried at 80 °C and then weighed to calculate the loss of polymer material.

*Differential Scanning Calorimetry (DSC)*

The phase transition of PNIPA-based hydrogels was investigated by Differential Scanning Calorimetry using a DSC Q200 from TA instrument. Gel samples (ca. 80 mg), equilibrated with a reference filled with the same quantity of solvent, were submitted to temperature cycles between 10 and 70 °C. The heating and cooling rates were fixed at 2 °C/min.

*Rheology*

The viscoelastic properties of hydrogels in their preparation state (discs with 4 cm diameter and 2 mm thickness) were studied using a stress-controlled rheometer (AR 1000 from TA Instruments) equipped with a plate/plate geometry (diameter 40 mm). The experiments were performed in the linear regime and the temperature was controlled by a high power Peltier system. The experimental conditions were fixed at constant frequency (1 Hz) and shear stress

(2 Pa) after applying a constant deformation of 20 % on the gel sample. A particular care was taken to avoid the drying of the sample by using a homemade cover which prevents from water evaporation during experiment. In these conditions, dynamic moduli ($G'$ and $G''$) were recorded between 20 and 70 °C by applying heating and cooling scans of 2 °C/min.

*Large strain behavior in tension mode*

Tensile tests were performed on a standard tensile Instron machine, model 5565, equipped with an environmental chamber allowing a precise control of the temperature. The device used a 10 N load cell (with a relative uncertainty of 0.16 % in the range from 0 to 0.1 N) and a video extensometer which follows the local displacement up to 120 mm (with a relative uncertainty of 0.1 % at full scale).

The gel samples used for mechanical tests were synthesized in home-made moulds consisting of two covered glass plates spaced by a stainless steel spacer of 2 mm thickness. The gels were then cut with a punch to their final dimensions: 30 mm x 4.9 mm x 2 mm. The gauge length was taken constant (L~20 mm) for all the tests and the gel strip was marked with two dots with a white marker, for their recognition by the video extensometer. For high temperatures or long-term experiments, tests were conducted in a temperature-controlled immersion cell consisting of a paraffin oil bath surrounding the sample in order to avoid water exchange with the environment. From swelling tests, carried out with dry PNIPA networks immersed during several days in paraffin oil at 60 °C, we have checked that this liquid is totally incompatible with PNIPA. Both monotonic tensile tests and fracture tests were performed in triplicate under controlled temperature applying a fixed strain rate of 0.06 s$^{-1}$. During the test, the force (F) and the displacement (L) were recorded while the nominal stress ($\sigma$=F/S$_0$) and the stretch ratio $\lambda$ ($\lambda$=L/L$_0$) were calculated. Fracture tests were carried out using the single edge notch geometry. A notch of approximately 1 mm length was made in the middle of a gel strip, whose total width was 5 mm and the length *L*~20 mm. The fracture energy, $G_c$ has been calculated using the following expression: $G_c=(6 \cdot W \cdot c)/\sqrt{\lambda_c}$ with $c$ the initial notch length, $\lambda_c$ the stretch ratio at break in single edge notch experiment and $W$ the strain energy density calculated by integration of the stress-strain curve.

*Small Angle Neutron Scattering (SANS)*

SANS experiments were performed at Laboratoire Léon Brillouin (CEA-Saclay, France) on the PAXY spectrometer. The wavelength of the incident neutron beam was set at $\lambda_0 = 12$ Å with a corresponding sample-to-detector distance of 4.7 m. This configuration provides a scattering vector modulus $q$ ranging between 0.003 and 0.05 Å$^{-1}$. For SANS experiments, gel plates of 2-mm thick were specially synthesized in D$_2$O to enhance the scattering contrast between the polymer network and the solvent.

For isotropic analyses, the gel discs (diameter = 14 mm and thickness = 2 mm) punched from plate samples in their preparation state were fit inside a ring spacer hermetically sandwiched between two quartz slides. The gel samples were then placed in a temperature controlled autosampler and let to equilibrate at least during 30 min at a given temperature (between 20 and 50 °C) prior to scattering experiments.

For anisotropic measurements performed with hydrogels under uni-axial deformation, a special device has been developed as described in previous papers.[20, 25] With this setup, the hydrogel strip, immersed into the thermostated chamber filled with perfluorodecalin, can be studied for hours without drying. In the following, all the anisotropic scattering experiments have been carried out during at least 1 hour for each sample in a given deformation state. For all the analyses, the efficiency of the detector cell was normalized by the intensity delivered by a pure water cell of 1-mm thickness and absolute measurements of the scattering intensity I($q$) (cm$^{-1}$ or 10$^{-8}$ Å$^{-1}$) were obtained from the direct determination of the incident neutron flux and the cell solid angle. Finally, the coherent scattering intensity of the gel was obtained after subtracting the contribution of the solvent used, as well as the perfluorodecalin for samples studied in the oil environment. For 2D SANS experiments, the incident neutron flux recorded on a two dimensional detector built up with 15500 cells of 25 mm$^2$, is directly transformed into a 2D image with a color code. After sector averaging, the data were quantitatively analyzed in terms of diffusion pattern.

## Results and discussion

*Transition temperature and swelling behavior*

Independently of their macromolecular structure, all PNIPA based hydrogels display a clear endotherm upon heating that corresponds to the overall energy balance of hydrogen bonds disruption/re-formation between amide groups and water molecules. Nevertheless, as shown in **Fig. 2a**, where integral curves (cumulated enthalpy) have been plotted versus temperature, some differences can be pointed out as a function of the gel architecture. We can first underline that the phase transition of hydrogels prepared with the same cross-linked PNIPA frame (GPN/PD and GPN-D), starts at the same association temperature, $T_{as}$, just above 30 °C. The transition enthalpy is nevertheless higher for the semi-IPN (4.3 kJ/mol NIPA) compared to the grafted PNIPA (2.9 kJ/mol NIPA) as the PDMA chains regularly grafted along the thermoresponsive backbone sterically hinder the aggregation process of NIPA units and their level of dehydration. By comparison, the two other hydrogels, tailored with a cross-linked PDMA frame, display much pronounced differences. As already described in previous papers,[20, 22] the phase transition of short PNIPA chains anchored on the chemically crosslinked PDMA network starts at higher temperature (32-33 °C). This does not just have to do with the low molar mass of PNIPA grafts but it is also related to the penalty of self-assembling PNIPA chains covalently attached at one end to the PDMA network. On the other hand, long PNIPA chains of semi-IPN GPD/PN, which were embedded into the PDMA network, start to self associate at lower temperature (~28 °C). The lower transition temperature of linear polymer PNIPA chains compared to PNIPA gels has already been reported in the literature and attributed to the higher apparent diffusion coefficient of linear chains relative to polymer network.[26] Qualitatively, we may assume that the higher transition enthalpy observed with semi-IPN architectures is accounting for a higher extent of the phase separation process; the less constraint linear PNIPA demonstrating larger enthalpy than the cross-linked PNIPA of opposite topology. On the other hand, grafted architectures have to compromise at a lower scale with a greater proportion of hydrated interface due to chemical bounds between PNIPA and PDMA domains.

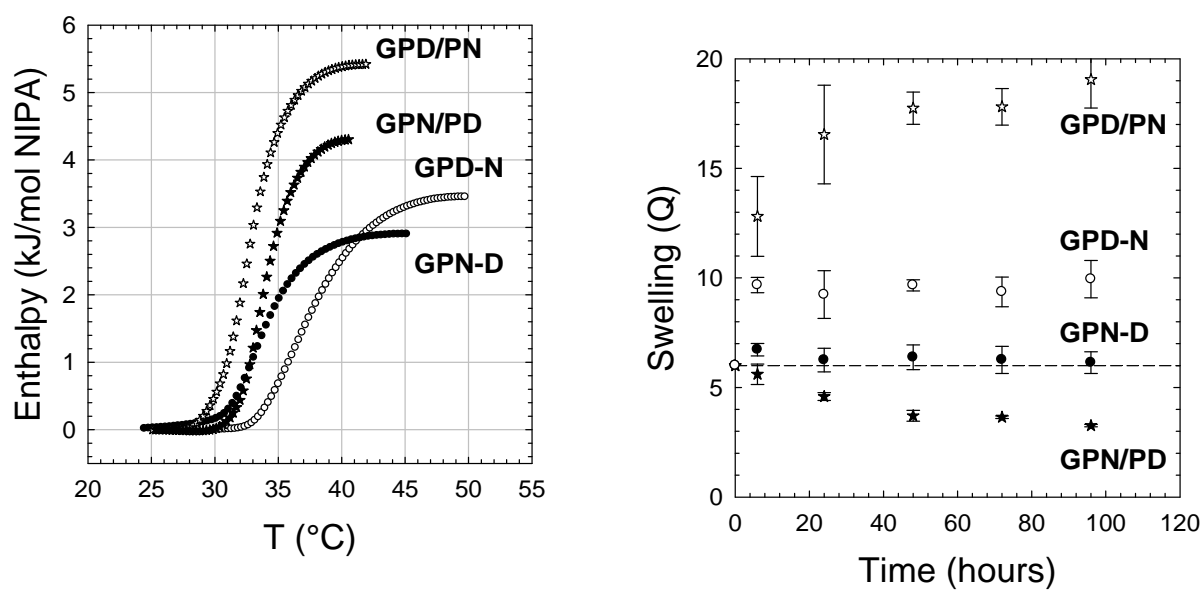

**Fig. 2**. a) DSC analysis of hydrogels at preparation state (heating rate=2 °C.min$^{-1}$);
b) Swelling kinetics of the same hydrogels at 60 °C in pure water.
GPD/PN (☆), GPN/PD (★), GPD-N (○) and GPN-D (●)

As demonstrated in former studies,[20, 22] the combination of thermo-responsive and hydrophilic polymers within the same network is crucial to control the level of swelling above the transition temperature. Indeed, as shown in **Fig. 2b**, grafted architectures prepared with equivalent weight fractions of PNIPA and PDMA are able to swell ($Q_e$=10 for GPD-N) or at least to sustain their initial swelling ($Q_0$=6 for GPN-D) when immersed at 60 °C in a large excess of water. Here the topology has a real impact on the volume transition as the crosslinked network mainly governs the swelling behavior. We effectively observe a higher shrinking ability of the gel when the thermoresponsive polymer is located in the cross-linked backbone (GPN-D) rather than in the side-chains (GPD-N). The situation is more complex in the case of semi-IPN networks as free linear chains, either PDMA or PNIPA, are able to diffuse inside and even outside the gel during long term swelling experiments. From analyses carried out after 4 days of equilibrium in water, the total polymer loss was about 5 wt% for GPD/PN and 20 wt% for GPN/PD. This means that when linear PNIPA chains phase separate within the hydrophilic network they are less prompt to diffuse out of the network (less than 10 % of PNIPA chains leave the gel during swelling). On the other hand, in the case of the GPN/PD network, about 40 % of hydrophilic PDMA chains can diffuse outside the shrinking network. Consequently, we observe slower kinetics for the two semi-IPN with a higher level of swelling for GPD/PN ($Q_e \cong 20$) and a much lower one for GPN/PD ($Q_e \cong 4$) after 4 days equilibrium at 60 °C. In the latter case, the deswelling from the preparation state is very progressive and for experiments performed on short time scale, typically below 4 hours, the collapse of the GPN/PD network remains low (< 15 %).

The main conclusion of these swelling experiments is that the mobility of linear chain in semi-IPN structures is responsible for polymer reorganization at larger scale above the transition temperature with the formation of PNIPA-rich domains alternating with the PDMA/water phase. The interesting feature is that, when gel samples are kept in their preparation state and isolated from water environment, they are able to retain their initial swelling ($Q_0=6$) and consequently their initial volume over the whole range of temperatures, i.e. well below and above the phase transition. These isochoric conditions, which have been effectively verified for GPD-N, GPN-D and GPD/PN which do not deswell below $Q_0=6$ at 60 °C, will be also assumed for GPN/PD that only slightly collapses on short time scales.

*Phase-separated structure at rest and linear viscoelasticity response*

SANS experiments were carried out with undeformed semi-IPN samples to investigate the phase transition taking place within the networks. As shown in **Fig. 3**, the phase separation of PNIPA starts at lower temperature (just above 28 °C) for GPD/PN compared to GPN/PD (T > 29-30 °C) as already described from DSC experiments. At high temperature, the scattered intensity of both hydrogels decays in the high-$q$ regime with the same Porod law ($I(q) \sim q^{-4}$) reporting the formation of a sharp interface between collapsed PNIPA domains and the PDMA/water phase. Contrary to grafted hydrogels, which display a clear correlation peak at high temperature (see **Fig. 4**), figuring fluctuation concentrations between rich- and poor-PNIPA phases with a characteristic distance ($d_c = 2\pi/q_{peak}$) of about 500 Å, the scattered intensity of semi-IPN increases continuously from large $q$ to low $q$ values underlining larger distances between bigger phase separated domains. As regards the PNIPA network grafted with long PDMA side-chains (GPN-$D_{long}$), it shows an intermediate behavior between grafted and semi-IPN architectures with a scattering peak shifted at lower $q$ values corresponding to a higher characteristic length of about 900 Å.

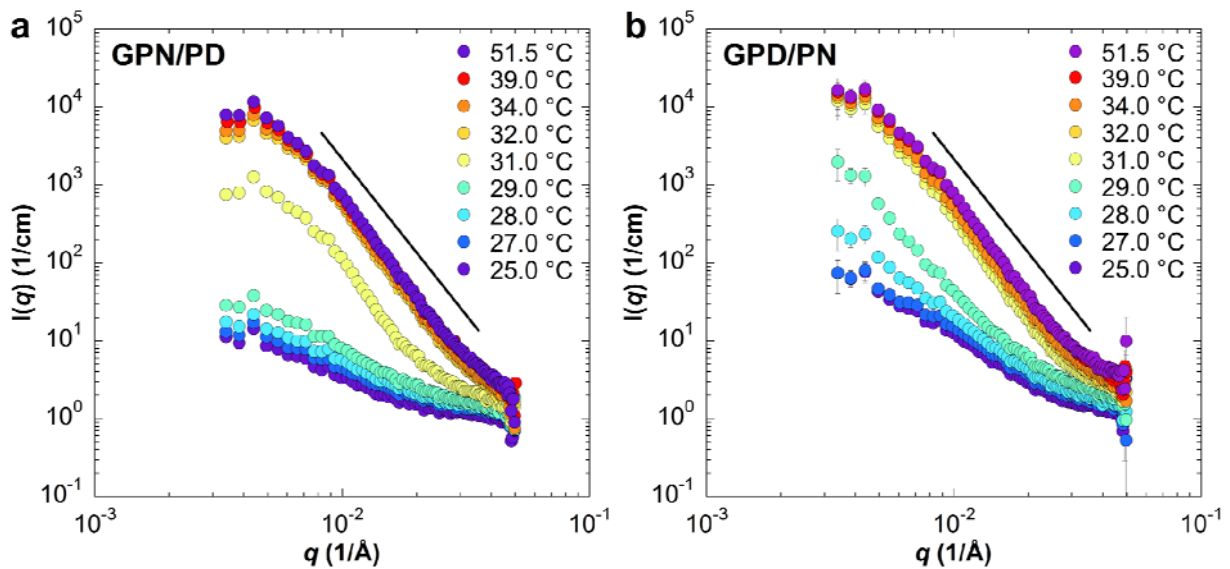

**Fig. 3.** Scattered intensity profiles of **(a)** GPN/PD and **(b)** GPD/PN semi-IPN hydrogels as a function of temperature. The solid lines indicate the Porod limit ($I \sim q^{-4}$).

Additional insights concerning the scale of the phase separation process can be obtained by calculating the total interface area of microdomains (*S*) in the scattering volume (*V*) using the asymptotic behavior obtained at high temperature ($I(q) \sim q^{-4}$) and the experimental Invariant $Inv = \int_0^\infty q^2 I(q) dq$, as follows :

$$S_{spe} \cong \frac{S}{V\phi_2\phi_1} = \frac{\pi}{Inv} \lim_{q \to \infty} q^4 I(q) \quad \{1\}$$

with $\phi_2$ the volume fraction of PNIPA domains and $\phi_1 = 1 - \phi_2$.

As $V\phi_2$ is the total volume of PNIPA microdomains and $\phi_1$ is close to 0.9 for all the gels in the segregated regime, the left hand term in equation {1} can be identified with the specific surface of PNIPA microdomains ($S_{spe}$).

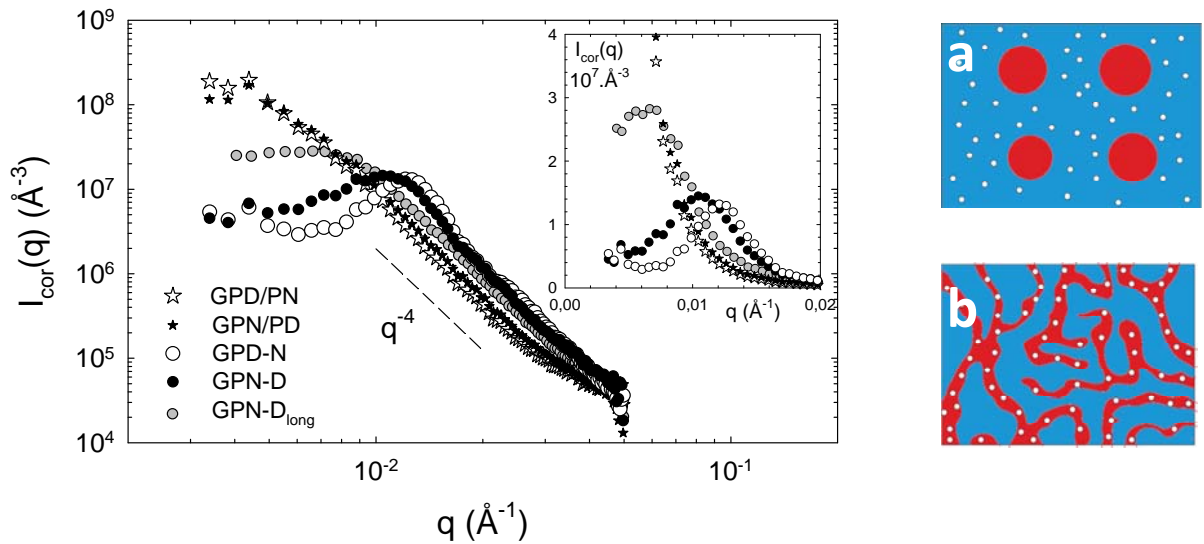

**Fig.4.** <u>Left</u>: Log-log plot of the normalized* scattered intensity profiles of GPD/PN (☆), GPN/PD (★), GPD-N (○), GPN-D (●) and GPN-D$_{long}$ (●) at 51.5 °C. The lin-lin plot is given in inset. <u>Right</u>: Micellar (**a**) and bicontinuous (**b**) microphase-separated morphologies with PNIPA-rich domains (red), PDMA swollen phase (blue) and chemical cross-linkers (white dots).

*SANS intensities were normalized by the Invariant ($Inv = 2\pi^2 (\Delta\rho)^2 \phi_2\phi_1$) in order to remove the contribution of the volume fractions of the two phases ($\phi_2$ the volume fraction of PNIPA-rich domains and $\phi_1=1-\phi_2$) as well as their contrast difference $(\Delta\rho)=\rho_2-\rho_1$. The normalized values ($I_{cor}(q) = 2\pi^2 I(q)/Inv$) were calculated using the experimental Invariant given below.

These values reported in **Table 2**, quantitatively point out the large differences between the phase separated morphologies taking place within semi-IPN and grafted architectures; the latter showing a much larger interface corresponding to smaller spacing of the microphase separated structures. From previous works, it has been observed that the biphasic morphologies obtained at high temperature with weakly crosslinked ([MBA]/[monomers]=0.1 mol%) and uncrosslinked grafted copolymers were mainly fixed by the nature, the size and the distribution of grafts along the main chain.[22, 24] Hydrophilic networks grafted with PNIPA side-chains (GPD-N) self-assemble at high temperature into a micellar morphology (see **Figure 4**), where the characteristic distance between hydrophobic domains is mainly controlled by the copolymer concentration, the solvent strength and the average molar mass of both hydrophilic spacers and thermoresponsive grafts.[27, 28] For the opposite topology, i.e. when the thermoresponsive backbone is grafted with hydrophilic side-chains (GPN-D networks), the macromolecular system self-assembles at high temperature into a bicontinuous morphology with a percolating cross-linked PNIPA phase (see **Figure 4**) which characteristics also depend on the primary structure of the copolymer[20, 22]. Indeed, as theoretically described by Borisov and coworkers for hydrophobic backbones grafted with hydrophilic side-chains,[29, 30] the axial dimension of macromolecular self-assemblies formed in dilute conditions is fixed 1) by the size of the hydrophobic core, which grows with the number of hydrophobic monomers within the spacer and the extent of the phase separation, and 2) by the span of crowded side-chains which increases with their molar mass. In the semi-dilute regime, as it is the case with GPN-D gels, the steric repulsions between PDMA coronas will also controls the characteristic length scale of the phase separation. Such behavior is highlighted with GPN-D networks in **Figure 4** and **Table 2**, where the specific surface of PNIPA domains decreases, i.e. the thickness of the rich-PNIPA phase as well as the characteristic length scale of the phase separation ($\xi \propto S_{spe}^{-1}$) increase, when the molar mass of PDMA side-chains and the average molar mass of PNIPA spacers increase ($M_{PDMA} \cong M_{PNIPA}$). In the case of semi-IPN networks, the characteristic length of the phase separation will depend on the molar mass of free polymer chains, their level of entanglement within the cross-linked network as well as the overall dynamics of the phase separation process. For GPN/PD, designed with the cross-linked PNIPA frame, one can expect a similar bicontinuous morphology than the one reported for grafted GPN-D at high temperature. Nevertheless in this case, the specific surface of the rich-PNIPA phase is much smaller underlining the formation of thicker PNIPA domains and larger characteristic distances between these domains. For the opposite semi-IPN topology (GPD/PN), the free PNIPA chains embedded into the hydrophilic cross-linked network self-assemble into larger aggregates as suggested by the lower specific surface of the rich-PNIPA phase. SANS spectra plotted in **Figure 3** do not really show great differences in this q-range between the biphasic morphologies adopted by the opposite semi-IPN networks. For this reason we will assume that GPD/PN adopt at high temperature an intermediate morphology with the formation of large and loosely connected PNIPA aggregates.

| Table 2. Specific surface of PNIPA domains at 51.5 °C ||
|---|---|
| Hydrogel | $S_{spe} \cong \dfrac{S}{V\phi_2\phi_1} = \dfrac{\pi}{Inv}\lim_{q\to\infty} q^4 I(q)$ |
| GPD/PN | 0.011 Å$^{-1}$ |
| GPN/PD | 0.013 Å$^{-1}$ |
| GPN-D$_{long}$ | 0.022 Å$^{-1}$ |
| GPN-D | 0.027 Å$^{-1}$ |
| GPD-N | 0.031 Å$^{-1}$ |

Complementary experiments performed in the preparation state by dynamic rheology demonstrate that the combination of PNIPA and PDMA chains into semi-IPN architectures enables to trigger thermo-hardening properties under isochoric conditions (**Figure 5**). Nevertheless, the mechanical hardening induced by the phase separation of PNIPA is much more efficient when PNIPA belongs to the network frame, that is the load bearing phase, rather than simply embedded as free chains that do not provide the same level of connectivity and dissipation above $T_{as}$. As previously reported with graft copolymer networks, the variation of the viscoelastic properties are totally reversible with temperature but the transition temperature remains dependent on the thickness of the gel sample and the scanning rate. Indeed, the symmetric hysteresis observed between heating and cooling cycles can be highly reduced from $\Delta T=30$ °C to 5 °C by decreasing the scanning rate from 10 to 0.5 °C/min, respectively.[22]

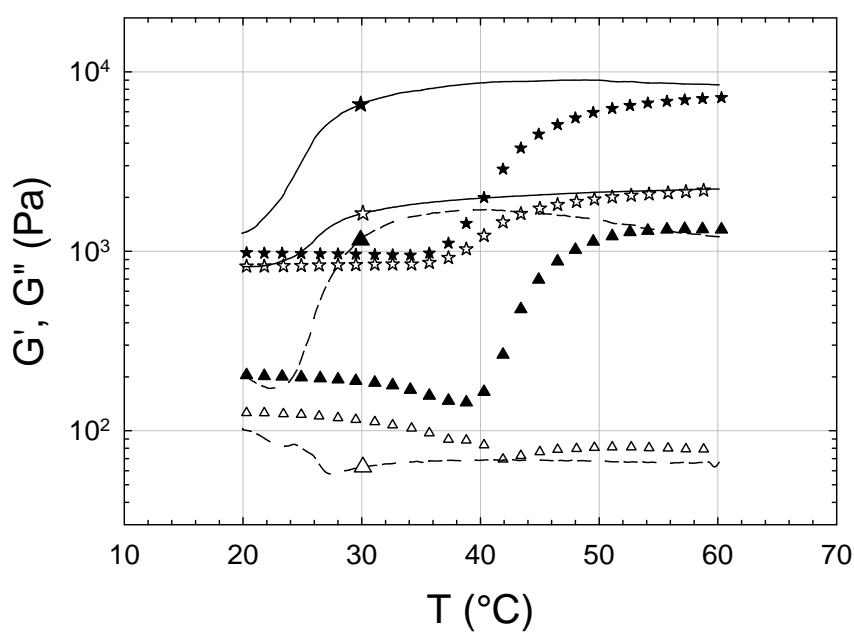

**Fig. 5.** Temperature dependence (during heating) of viscoelastic properties at 1 Hz of GPN/PD (★, ▲) and GPD/PN (☆, △) hydrogels at preparation state applying a heating rate of 2 °C/min (stars symbols stand for $G'$ and triangle symbols for $G''$). The solid and short dash lines correspond, respectively, to $G'$ and $G''$ during subsequent cooling at 2 °C/min.

The comparison between elastic moduli of all hydrogels plotted in **Figure 6,** clearly points out the close behavior of GPN-D, GPN-D$_{long}$ and GPN/PD at low deformation.

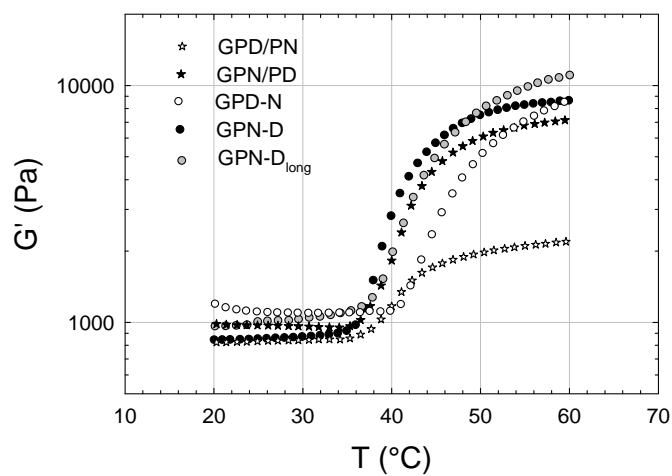

**Fig. 6.** Temperature dependence of the elastic modulus (*G'*) at 1 Hz of GPN/PD (★), GPD/PN (☆), GPN-D (●), GPN-D$_{long}$ (◐) and GPD-N (○) hydrogels at preparation state applying a heating rate of 2 °C/min.

In this case the scale of the phase separation does not seem to matter on the viscoelastic properties at low deformation and the same conclusion holds for the comparison between GPN-D and GPN-D$_{long}$ prepared with short and long PDMA grafts (M$_n$=39 and 89 kg/mol, respectively). If we assume that the volume fraction of PNIPA-rich domains ($\phi_2$) as well as the PNIPA weight fraction inside these domains are similar in all hydrogels prepared with a crosslinked PNIPA frame, this mechanical behavior can be interestingly compared with open-cell polymer foams where the elastic modulus of the material varies only with the modulus of the wall and the square of their volume fraction.[31] By comparison, the sharp phase separation that takes place within the GPD/PN hydrogel does not allow the formation of a PNIPA phase able to sustain the mechanical stress. Moreover, as there are no chemical connections between PNIPA and PDMA phases in GPD/PN, its properties are even much lower than GPD-N which micellar morphology formed at high temperature gives rise to an efficient coupling between the covalent crosslinks of the PDMA matrix and additional physical crosslinks formed by collapsed PNIPA microdomains.

*Large strain behavior: structure and mechanical properties*

The mechanical properties of GPD/PN and GPN/PD hydrogels reported in **Fig.7** have been carried out at 20 and 60 °C with a strain rate of 0.06 s$^{-1}$.

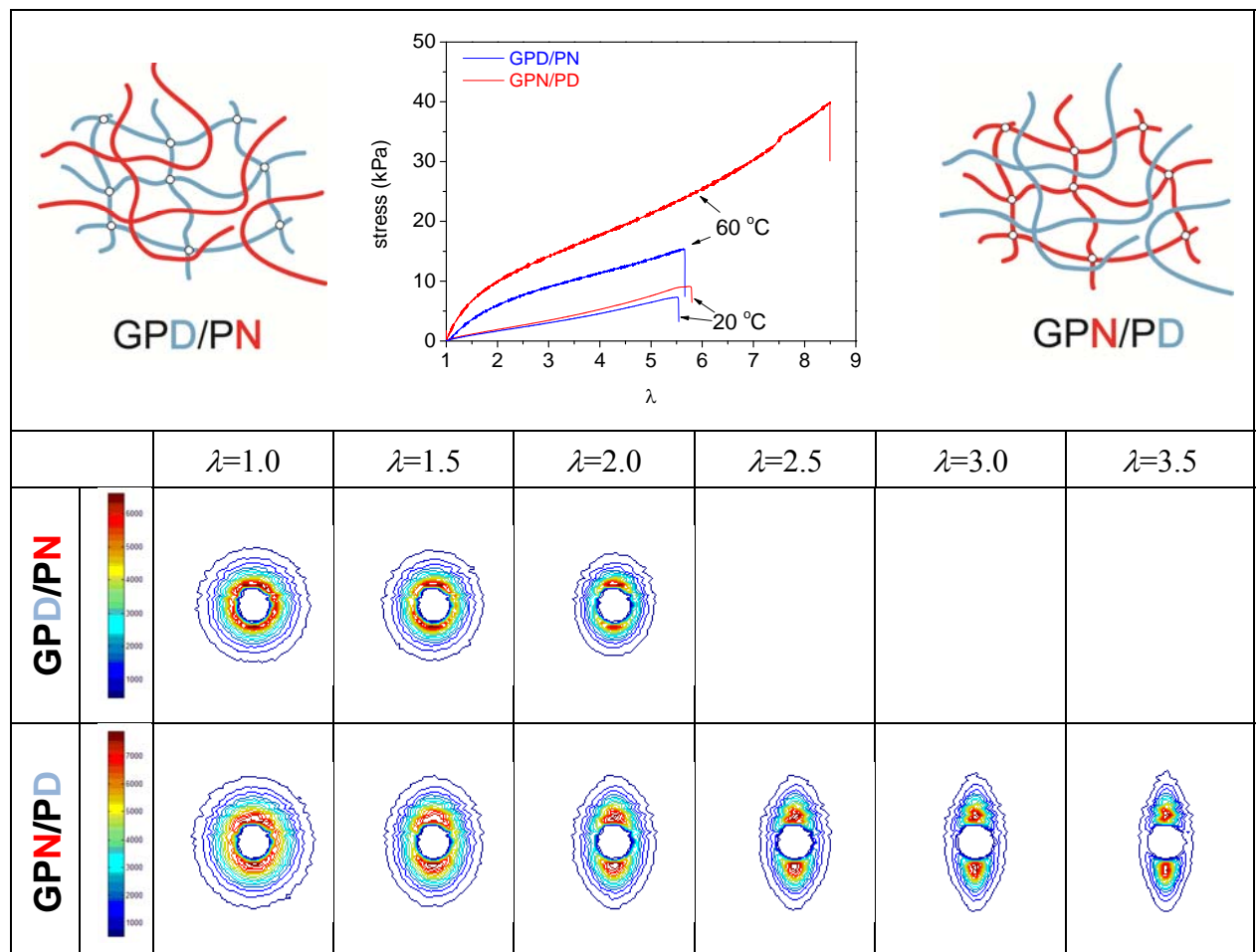

**Fig. 7.** Structure and properties of semi-IPN GPD/PN and GPN/PD under deformation. Primary structure of GPD/PN and GPN/PD hydrogels and their uniaxial tensile stress-strain curves at 20 and 60 °C. Iso-intensity SANS patterns were obtained after a step-by-step loading at 60 °C. The direction of the uniaxial deformation is along the horizontal axis.

For these mechanical tests, performed in an oil bath at 60 °C, we can reasonably assume that the gels keep their initial dimensions. This is obvious for GPD/PN, as PNIPA chains cannot diffuse in the oil phase which is totally incompatible and water molecules remain in the network because of the positive osmotic pressure. For GPN/PD, the same holds for linear PDMA chains which remain confined into the PDMA network. Only water can be expelled at the surface of the gel when heated at 60 °C but as the volume shrinkage is relatively slow compared to the duration of the mechanical test (10 minutes for thermal equilibrium and less than 2 minutes for the tensile test), this phenomenon will be considered as negligible in these conditions.

At 20 °C, when both PNIPA and PDMA chains are water-soluble, both gels exhibit a very similar mechanical behavior. On the other hand, working at 60 °C well above the critical temperature of PNIPA, the phase separated semi-IPN GPN/PD demonstrates a much larger mechanical reinforcement than GPD/PN with larger increase of both tensile modulus and extension at break. The work of extension, defined by the area under the tensile curve, is then more than 4 times higher for GPN/PD (190 kJ.m$^{-3}$) compared to GPD/PN (42 kJ.m$^{-3}$), confirming that the thermo-reinforcement of the mechanical properties is much more efficient when the responsive polymer is forming the crosslinked frame of the hydrogel.

As PNIPA-rich domains are forming quenched heterogeneities at high temperature, these scatterers can be used as probes to follow their average displacement under mechanical deformation and to investigate the local structure of thermo-responsive hydrogels. For that purpose, a series of SANS experiments have been carried out on GPN/PD and GPD/PN hydrogels, under incremental stretch ratio, from $\lambda$ = 1 to 2-3.5 at 60 °C, i.e. well above the phase transition temperature. The 2D SANS spectra, reported in **Fig. 7**, show that both undeformed samples ($\lambda$ = 1) display at 60 °C an isotropic pattern with a circular diffraction ring and a maximum of intensity localized close to the beam stop as already described with static experiments in **Fig. 3**. Under deformation, the 2D spectra are distorted into prolate ellipsoid with their major axis along the perpendicular direction due to the orientation and deformation of PNIPA-rich domains. Due to the nature of reciprocal space, this means that the frequency with which the neutron density difference is encountered is greater perpendicular to the direction of orientation than along the direction of orientation.

At the same stretch ratio, the distortion of the 2D pattern is much higher for GPN/PD highlighting larger local deformation of the PNIPA phase within this topology. Its behavior is very similar to the one reported for the grafted hydrogel GPN-D with the same PNIPA backbone where an elliptically shaped correlation band with a non-uniform azimuthal intensity distribution was observed with two diffraction spots in the perpendicular direction.[22] The anisotropy induced upon stretching can be analyzed more quantitatively by plotting the parallel ($I_{//}(q)$) and perpendicular ($I_\perp(q)$) scattering intensities obtained after radially averaging the 2D data within a rectangular sector of axis, respectively, parallel (//) or perpendicular ($\perp$) to the deformation axis (**Fig.8**).

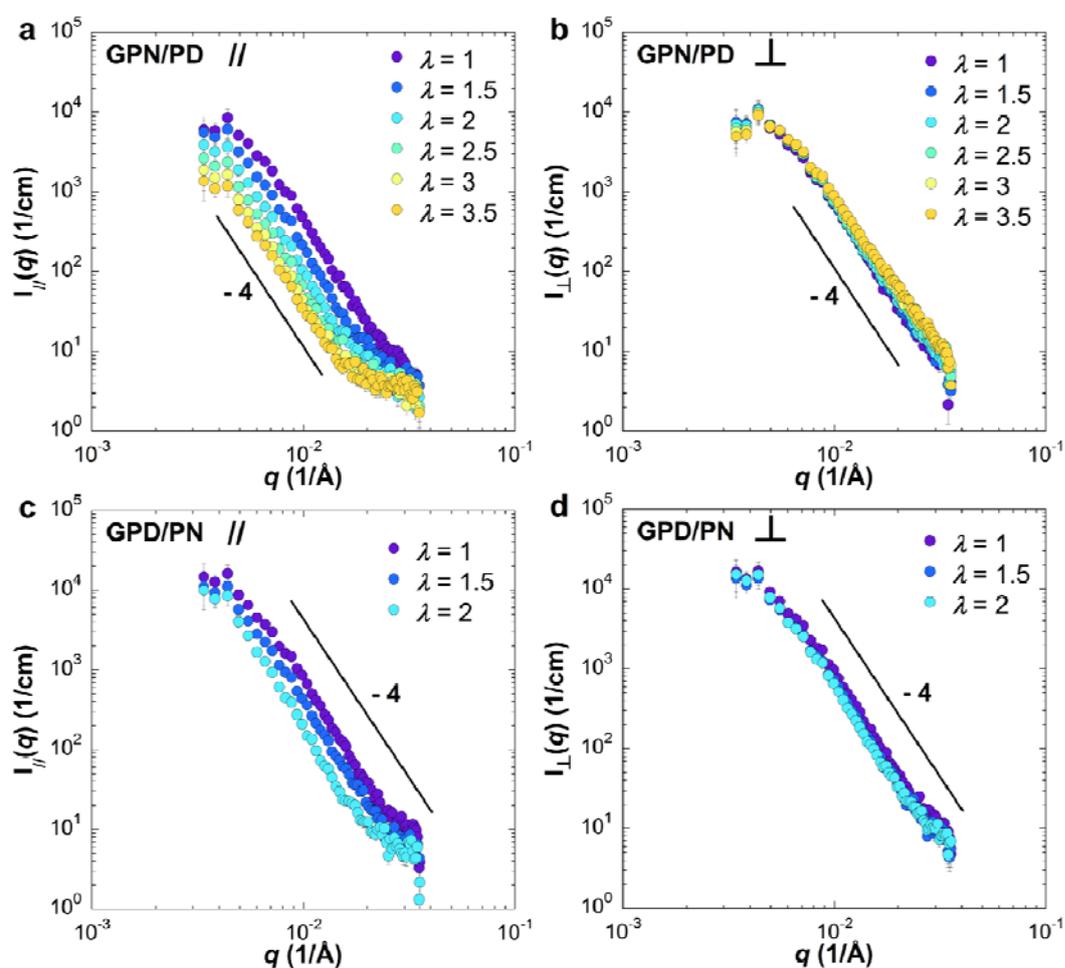

**Fig.8**. Scattered intensity profiles at 60 °C of semi-IPN gels under uni-axial extension. **(a, b)** GPN/PD (from $\lambda = 1$ to $\lambda = 3.5$) and **(c, d)** GPD/PN (from $\lambda = 1$ to $\lambda = 2$). The data were obtained from 2D-SANS after sector averaging in the parallel (//) and perpendicular ($\perp$) directions of the deformation axis.

In the absence of characteristic distance observable in the *q*-range investigated the main characteristics of the two semi-IPNs at high temperature are the very weak and very strong dependences of the scattered intensity versus the deformation along the perpendicular and equatorial axes, respectively.

As suggested by Effler et al.,[32] a simple characterization of the orientation undergone at the local scale by the sample would be to measure the ratio of Invariants determined along the stretching and transverse directions ($Inv_{//}/Inv_{\perp}$). As shown in **Fig.9**, the anisotropic ratio $Inv_{//}/Inv_{\perp}$ strongly decreases with increasing strain. This means that the rich-PNIPA phase, which was initially randomly oriented within the network, becomes rapidly oriented in the direction of the deformation as schematically shown in **Figures 9a** and **9b**. For a given stretch ratio, the larger orientation of the semi-IPN GPN/PD compared to its opposite topology GPD/PN, can be attributed to a greater connectivity of the rich-PNIPA phase as already suggested from rheology and tensile tetts.

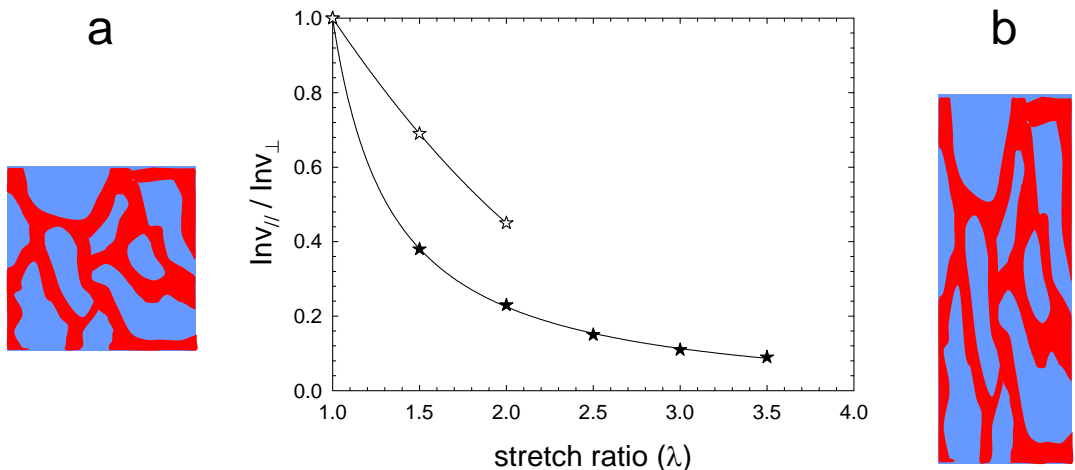

**Fig. 9**. Strain dependence of the Invariant ratio between parallel ($Inv_{//}$) and perpendicular ($Inv_{\perp}$) components for GPD/PN (☆) and GPN/PD (★) hydrogels.
**(a)** and **(b)** are schematic representations of the biphasic network (with rich-PNIPA phase in red) at rest and under uniaxial stretching, respectively.

While SANS and tensile tests clearly emphasize the large impact of the primary structure on both self-assembling and toughness above $T_{as}$, the difference between the two gels becomes even more significant when comparing the fracture tests performed on notched samples. As shown in **Fig. 10**, the GPN/PD gel with PNIPA forming the cross-linked backbone can reach fracture energy as high as 1.5 kJ/m² at 60 °C with a systematic crack bifurcation. This energy is even much higher than the graft counterpart GPN-D$_{long}$ grafted with long pendant PDMA chains. As already emphasized with GPN-D samples designed with various grafting densities,[20] the comparison with the semi-IPN GPN/PD confirms the close relationship between the cross-linked architecture, the scale of the phase separation and the mechanical properties. As pictured in **Fig. 10**, the thicker is the PNIPA phase and the higher will be the fracture resistance with original crack bifurcation specifically obtained in the case of GPN-D and GPN/PD networks.

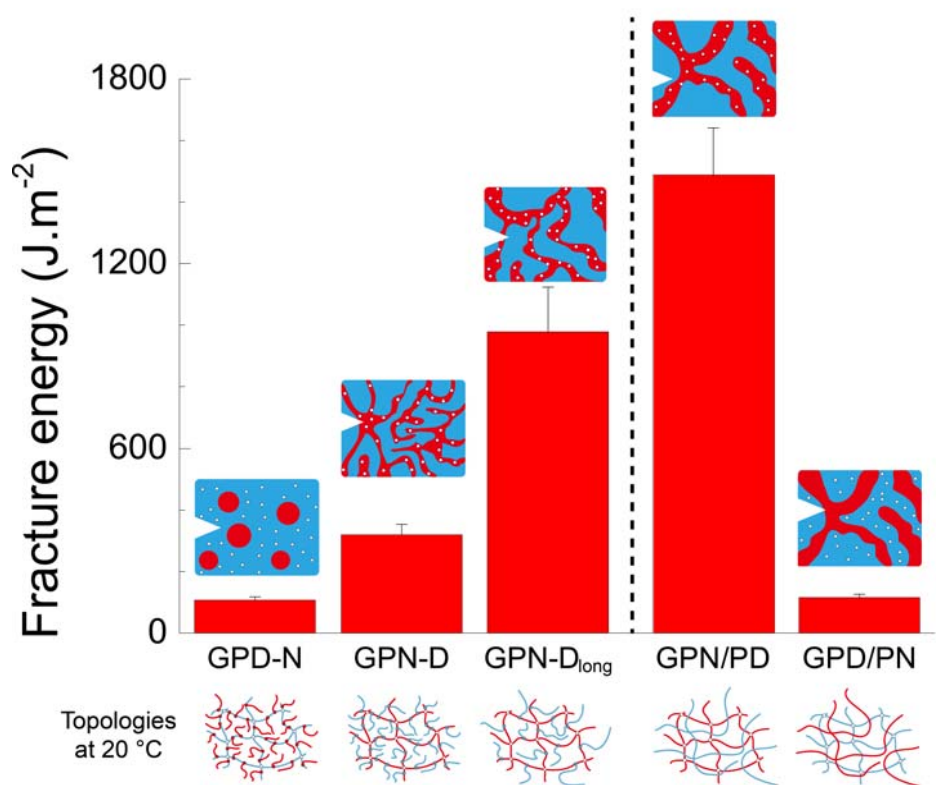

**Fig. 10**. Fracture energies obtained with PNIPA/PDMA hydrogels at 60 °C and schematic representation of the morphologies developed at high temperatures with PNIPA-rich domains (red), PDMA swollen phase (blue) and chemical cross-linkers (white dots).

By contrast, the other semi-IPN GPD/PN can only achieved fracture resistance of one decade lower than GPN/PD; typically at the same level as the GPD-N hydrogel although they display different intrinsic morphology at high temperature. For the semi-IPN GPD/PN, we assume that the phase separation of linear PNIPA chains gives rise to weakly connected and uncrosslinked domains that cannot easily withstand the stress and prevent the crack propagation. The situation is different for the grafted GPD-N network where the self-assembling of PNIPA grafts at 60 °C into isolated micellar domains effectively contributes to the improvement of the elastic behavior (increase of the modulus and work of extension) but cannot prevent efficiently the crack propagation through the PDMA matrix.

## Conclusion

Within the development of thermo-toughening hydrogels, the aim of this work was to combine hydrophilic chains (PDMA) and LCST polymers (PNIPA) in semi-interpenetrated architectures in order to investigate their responsive properties in swollen isochoric conditions comparatively to grafted network structures. By keeping constant and equal the weight fractions of PDMA and PNIPA in hydrogels, the new semi-IPN have both demonstrated thermoresponsive behavior with very different structure/properties relationships. In the case of the PDMA network interpenetrated by linear PNIPA chains (GPD/PN), the phase transition of PNIPA occurs at lower temperature comparatively to other hydrogels and gives rise to large microdomains with a very weak percolation through the PDMA network. The enhancement of the mechanical properties above the transition temperature is indeed very low with a 3-times increase of the elastic modulus and little improvement of the fracture energy; the isolated PNIPA phase acting mainly like a filler in polymer materials.

Conversely, the interpenetration of PDMA chains within the PNIPA crosslinked network (GPN/PD) brings a large improvement of the mechanical properties at high temperature with a 10-fold increase of the modulus and a very high fracture energy. Nevertheless, the absence of connectivity between PNIPA and PDMA phase is responsible for the large scale of the phase separation which induces a small collapse of the network at 60 °C when starting from the preparation state.

From a general point of view, the comparison between semi-IPN and grafted networks clearly emphasizes the role of the primary structure over the phase-separated morphology and the resulting mechanical properties. It is obvious that the best reinforcement can be achieved by designing the gel with a PNIPA crosslinked frame and by implementing hydrophilic polymers in order to maintain a high level of swelling in the segregation regime. The comparison between GPN-D hydrogels brings highlights that the molar mass of PNIPA between hydrophilic side-chains is an important parameter for controlling the length scale of the phase separation; the larger being the best in the case of GPN-D versus GPN-D$_{long}$. Nevertheless, in absence of chemical connection between PDMA and PNIPA, as in GPN/PD, the length scale of the phase separation becomes even larger leading at the same time to a very good mechanical reinforcement but to a lower control of the gel volume above the transition temperature. There is consequently an optimum in the structure/properties relationships that will depend on the required properties above the transition. This could be found by exploring different alternatives to control the level of the phase separation, playing with the chemistry of the precursors, with the environmental conditions (temperature, ionic strength, pH,…) and using various architectures like grafted, multiblock or even fully interpenetrated networks that will allow the percolation of the PNIPA phase and the control of the segregation length scale with the degree of network crosslinking.

# Acknowledgements

We gratefully acknowledge the financial support of CNRS, ESPCI, and UPMC and the China Scholarship Council for the Ph.D. fellowship funding of H.G. The authors also thank Guylaine Ducouret from SIMM and Annie Brûlet from LLB (CEA, Saclay) for their assistance on rheology and small angle neutron scattering, respectively.